\documentclass[a4paper,14pt, ]{article}

\usepackage{mathrsfs}
\usepackage{cmap}					
\usepackage{mathtext} 				
\usepackage[T2A]{fontenc}			
\usepackage[utf8]{inputenc}			
\usepackage[english]{babel}	
\DeclareUnicodeCharacter{00A0}{}
\usepackage{braket}
\usepackage{csquotes}
\usepackage{amsmath,amsfonts,amssymb,amsthm,mathtools} 
\usepackage{icomma} 
\usepackage{pb-diagram}

\usepackage{graphicx}  
\graphicspath{{images/}{images2/}}  
\usepackage{wrapfig} 

\usepackage{array,tabularx,tabulary,booktabs} 
\usepackage{longtable}  
\usepackage{multirow} 

\theoremstyle{plain} 

\theoremstyle{definition}

\theoremstyle{remark} 

\usepackage{extsizes} 
\usepackage{geometry} 
\usepackage{tikz}
\usepackage{fancyhdr} 
\usepackage{setspace} 
\usepackage{titlesec}

\usepackage{lastpage} 

\usepackage{hyperref}
\usepackage[usenames,dvipsnames,svgnames,table,rgb]{}
\hypersetup{				
	unicode=true,           
	pdftitle={?????????},   
	pdfauthor={?????},      
	pdfsubject={????},      
	pdfcreator={?????????}, 
	pdfproducer={?????????????}, 
	pdfkeywords={keyword1} {key2} {key3}, 
	colorlinks=true,       	
	linkcolor=red,          
	citecolor=green,        
	filecolor=magenta,      
	urlcolor=cyan           
}

\begin{document}

\title{Quantum-like modeling of the order effect in decision making:  POVM viewpoint on the Wang-Busemeyer QQ-equality}

\author{Aleksandr Lebedev and Andrei Khrennikov\\
International Center for Mathematical Modeling \\
in Physics and Cognitive Sciences\\ 
Linnaeus University, V\"axj\"o, SE-351 95, Sweden}
\date{}
\maketitle

\begin{abstract}
In recent years, quantum mechanics has been actively used in areas outside of physics, such as psychology, sociology, theory of decision-making, game theory, and others. In particular, quantum mechanics is used to explain the paradoxes arising in cognitive psychology and decision making. Wang and Busemeyer  invented a quantum model and approach as well as non-parametric equality (so-called QQ-equality), 
explaining the questions order effect. The primary objective of this note is to test the possibility to  expand the Wang-Busemeyer model  by considering questions which are mathematically represented  by positive operator valued measures.  We found that, for such observables, the QQ-equality can be violated. But, we also showed that, in principle, it is possible to reduce expanded model to the original Wang-Busemeyer model by expanding the context of the questions. This version of preprint is aimed to point out to annoying miscalculation in version 1. This miscalculation might mislead a reader who is not experienced in operating with POVMs. Otherwise the main line of construction and reasoning presented in version 1 is right and it can be 
easily completed by the reader on the basis of version 1 and the correction remark in version 2. 
\end{abstract}

\section{Introduction}

This version is aimed to point out to annoying miscalculation in version 1  \cite{L}. 

Quantum mechanics was originally created to explain the paradoxes  arising in  classical physics. 
At the same time, a powerful mathematical apparatus of quantum probability theory was created, which was later effectively used 
to explain the paradoxes not only in physics, but also in other fields, such as cognitive psychology, decision making, and social 
sciences, see, for example, monographs \cite{ID} - \cite{ID5} and a few recent representative papers  \cite{ID6} - \cite{ID16}.
In particular, Wang and Busemeyer \cite{WB}used the quantum formalism and methodology of experiment to explain order effects in question answering.

Wang and Busemeyer established a non-parametric inequality (known as QQ-inequality)  to which the probabilities 
of an experiment must satisfy in order for a quantum model to exist for them, as follows:
$$p(AyBn)+p(AnBy)-p(ByAn)-p(BnAy)=0,$$
where $A$ and $B$ correspond to questions with two possible 
outcomes \enquote{Yes} and \enquote{No}.  The joint  probabilities are the probabilities of receiving given answers to questions $A$ and $B$ in the same order as they appear, 
e.g. $P(AyBn)$ means the probability to obtain negative answer to the question $B$ before obtaining affirmative answer to the question $A.$
The quantum-like model assumes that questions are  represented by Hermitian operators; therefore the answers 
$Ay, An, By,  Bn$ are represented by  orthogonal projectors.

The  following questions naturally arise:
\begin{enumerate}
\item Is it imperative to require these operators to be projectors?
\item Is it possible to expand the context  of the questions in such a way that, although the original operators are not projectors, 
the extended questions would already correspond to the projectors?
\end{enumerate}

The paper presents  examples of measurement operators corresponding to POVM for which the QQ-inequality does not hold.
The dependence  of the left part of the QQ-inequality on the state is discussed. 
Further, using the  Naimark's theorem \cite{pr}, lifting of such operators are constructed for which this equality does hold.

	\section{The example of violation QQ-inequality for POVM-observables}
	\label{QQV}

POVM stands for 'Positive operator-valued measure'. More precisely, it is the set of measurement operators $\{ E_a \}$ (called effects)  that form a complete set of Hermitian non-negative operators. It means that it has the following properties:
\begin{enumerate}
\item $E_a=E^{\dag};$	
\item $\bra{\phi} E_a\ket{\phi}\ge 0$  for all vectors $\ket{\phi}$
\item $ \sum_a E_a = 1$
\end{enumerate}	
	
 For POVMs represented as 
\begin{equation}
\label{LO99} 
E_a=M_a M^{\dag}_a,
\end{equation} 
 the state update generated by  a measurement is given by the formula
\begin{equation}
\label{LO} 
 \rho \to  \rho_a= \frac{M_a \rho M^{\dag}_a}{tr(E_a\rho)}.
\end{equation} 
In particular, for a pure state $|\psi\rangle,$ 
\begin{equation}
\label{LO0} 
 |\psi\rangle  \to  |\psi_a \rangle= \frac{M_a |\psi\rangle }{||M_a |\psi\rangle|| }.
\end{equation} 

We remark that if $M_a$ is projection, then the effect $E_a$ is also projection. Such POVMs are called projection observables. 
The inverse is not correct. An effect $E_a$ that is projection can be decomposed with operators $M_a$ (see {(\ref{LO99})) 
that are not projections. 

In particular, POVM does not determine the state update operation.  The right theory for the description of observations and 
the corresponding state updates is quantum instrument theory (see \cite{OJMP} for its simple presentation). However, in this note we do not apply this theory and we proceed with formulas (\ref{LO99}), (\ref{LO}). Within quantum instrument theory, these formulas determine 
so called atomic instruments.

	The operators corresponding to two measurement procedures are 
\begin{equation}
\label{LO1} 
	An + Ay=I,  Bn + By=I,
\end{equation}
	where labels $n,y$ denote the answers ``no'' and ``yes''. 
For the first equality, we select the operators: 
	\begin{equation}
\label{LO2} 
	\begin{pmatrix} 5/6 && 1/2 \sqrt{3} \\ 1/2 \sqrt{3} && 1/2 \end{pmatrix} + \begin{pmatrix} 1/6 && -1/2 \sqrt{3} \\ -1/2 \sqrt{3} && 1/2 \end{pmatrix} = I.
	\end{equation}
	Moreover, the second equality, we select the operators: 
\begin{equation}
\label{LO3} 
\begin{pmatrix} 1/6 && 1/2 \sqrt{3} \\ 1/2 \sqrt{3} && 1/2 \end{pmatrix} + \begin{pmatrix} 5/6 && -1/2 \sqrt{3} \\ -1/2 \sqrt{3} && 1/2 \end{pmatrix} = I.
\end{equation}
	We need to calculate the following quantity:
\begin{equation}
\label{LO4} 
qq =p(AyBn)+p(AnBy)-p(ByAn)-p(BnAy).
\end{equation}
	where $A$ represents the first measurement and $B$ represents the second one and vice verse.
	
	For the update, we should use, e.g., {\it square roots} from the (Hermitian) operators $An,Ay, Bn,By$ (cf. version 1 \cite{L} in that we misleadingly used not square roots, but the question-operators),
	\begin{equation}
\label{LO5} 
	Q_1= \sqrt{An}, Q_2= \sqrt{Ay}, \; P_1= \sqrt{Bn}, P_2= \sqrt{By}.
\end{equation}
In our terms,  (\ref{LO4}) has the form:
	$$qq(\rho) = Tr(P_2Q_1\rho Q_1P_2+P_1Q_2\rho Q_2P_1-Q_2P_1\rho P_1Q_2-Q_1P_2\rho P_2Q_1)=$$
	$$=Tr(P_2Q_1\rho Q_1P_2+P_1Q_2\rho Q_2-Q_2P_1\rho P_1-Q_1P_2\rho P_2Q_1)$$
	$$=Tr( Q_1P_2P_2Q_1\rho+Q_2P_1Q_2\rho -P_1Q_2P_1\rho -P_2Q_1Q_1P_2\rho ). 
	$$
It is easy to check that e.g. for the pure state $\psi= (1,1)/\sqrt{2},$
$$
qq(\rho_\psi) \not=0,
$$
where $\rho_\psi= |\psi\rangle \langle \psi|.$ 

This is the good place to emphasize that the Wang-Bussemeyer's QQ-effect is state dependent. This is not just the property of questions, but questions asked to people in the special mental states. The state dependence plays the crucial role in quantum-like modeling. Recently its role was highlighted in article \cite{QLO} advertising quantum logic of human mind. 
	
\section{Lifting of POVMs to projections:  Naimark's dilation theorem }

 Naimark's theorem states that, for any given POVM,  the Hilbert space,
 can be extended to a larger space that POVM can be realized as performing orthogonal measurements in that larger Hilbert space. We denote this process of obtaining the mapping from POVM to orthogonal measurements as 'lifting'.
	We can now consider 3-dimensional space and new operators
	\begin{equation}
\label{LO1a} 
	A^\prime n + A^\prime y=I,  B^\prime n + B^\prime y =I.
\end{equation}
For the first equality, we select the operators: 
	$$\begin{pmatrix}
	\frac{5}{6} && \frac{1}{2\sqrt{3}} && \frac{1}{3\sqrt{2}} \\
	\frac{1}{2\sqrt{3}}  && \frac{1}{2} && -\frac{1}{\sqrt{6}} \\
	   \frac{1}{3\sqrt{2}} && - \frac{1}{\sqrt{6}} && \frac{2}{3} 
	\end{pmatrix}+\begin{pmatrix}
	\frac{1}{\sqrt{6}} && -\frac{1}{2\sqrt{3}} && -\frac{1}{3\sqrt{2}} \\
	-\frac{1}{2\sqrt{3}}  && \frac{1}{2} && \frac{1}{\sqrt{6}} \\
	-\frac{1}{3\sqrt{2}} &&  \frac{1}{\sqrt{6}} && \frac{1}{3} 
	\end{pmatrix}=1$$
For the second equality, we select the operators: 
	$$\begin{pmatrix}
	\frac{1}{\sqrt{6}} && \frac{1}{2\sqrt{3}} && -\frac{1}{3\sqrt{2}} \\
	\frac{1}{2\sqrt{3}}  && \frac{1}{2} && -\frac{1}{\sqrt{6}} \\
	-\frac{1}{3\sqrt{2}} &&  -\frac{1}{\sqrt{6}} && \frac{1}{3} 
	\end{pmatrix}+\begin{pmatrix}
	\frac{5}{6} && -\frac{1}{2\sqrt{3}} && \frac{1}{3\sqrt{2}} \\
		-\frac{1}{2\sqrt{3}}  && \frac{1}{2} && \frac{1}{\sqrt{6}} \\
		\frac{1}{3\sqrt{2}} && \frac{1}{\sqrt{6}} && \frac{2}{3} 
	\end{pmatrix}=1$$
	Note that these operators are self-adjoint. Furthermore, they are projectors.
They are lifting of the operators considered in section \ref{QQV}
 in equations (\ref{LO2}), (\ref{LO3}).	Since now questions are mathematically represented by projections, the QQ-equality for them holds true for any state. 

\subsection{Interpretation of POVM-questions  and experimenting }

POVMs of non-projection type represent fuzzy measurements. ``Fuzziness'' is in the procedure of the quantum state update. If in (\ref{LO}) the operator $M_a$ is not projection, then generally, for a pure state $|\psi\rangle,$  updated states $|\psi_a\rangle, a=n,y,$ 
are not orthogonal. Alice by answering ``no'' to say question $A$ does not wash out the possibility to answer ``yes'', if the 
same question will be asked immediately.  Questions represented by projection observables are characterized by repeatability of answers, so by answering ``no'' to the $A$-question, Alice will answer again ``no'' to this question with probability one.

This difference between projection and non-projection questions can be tested experimentally. The indirect test is based on testing the combination of question order and response replicability effects \cite{OJMP}. However, it would be interesting to perform direct testing. 

Another really challenging problem is to check experimentally the consequences of the Naimark's dilation theorem. To find a question $A$ that cannot be 
represented by projective measurement, but can be represented by an atomic POVM $A.$ Then lift $A$ to a higher dimensional state space as  projection observable $A^\prime$ and realize the latter as a question completing the original question $A.$  
	
	\section{Acknowledgment} The authors would like to thank Jerome Bussemeyer who pointed out to the aforementioned miscalculation in version 1 and stimulated us to submit this  version.

\end{document}